# An interactive engine for multilingual video browsing using semantic content

M. BEN HALIMA, M. HAMROUN, S. BEN MOUSSA and A. M. ALIMI
REGIM-Lab: REsearch Group on Intelligent Machines *(*University of Sfax, National School of Engineers*)*
B.P 1173. 3038 Sfax-Tunisia

mohamed.benhlima@ieee.org, mohamed.hamroun@gmail.com, sami.benmoussa@gmail.com, adel.alimi@ieee.org.

**Abstract.** The amount of audio-visual information has increased dramatically with the advent of High Speed Internet. Furthermore, technological advances in recent years in the field of information technology, have simplified the use of video data in various fields by the general public. This made it possible to store large collections of video documents into computer systems. To enable efficient use of these collections, it is necessary to develop tools to facilitate access to these documents and handling them. In this paper we propose a method for indexing and retrieval of video sequences in a video database of large dimension, based on a weighting technique to calculate the degree of membership of a concept in a video also a structuring of the data of the audio-visual (context / concept / video) and a relevance feedback mechanism.

## 1. Introduction

In this work we aim to propose a method for video search that puts the user at the center of interest. The involvement of the latter in the research process reveals his needs and desires. Our contribution consists, first, in order to consolidate and organize data to facilitate data mining in research:

- Three possible levels of structuring data: contextual level, conceptual level and the level of raw data.
- Weighting concepts, using a well-defined formula.

Second, the relevance feedback technique proposed in the search query text to address the real needs of the user. To achieve our goal we track the following steps:

- Assist the user to reformulate his query based on the concepts. The renovation is evident in how the co-determine some concepts from words entered by the user by projection on a well-defined ontology.
- Communicate with the user in the display result; the work of relevance feedback is made in the context of text and image but very little work focused on video. We propose an algorithm of information retrieval in video documents based concepts.

Finally, to compute the similarity between the shot query and those of the collection we use a hybrid approach which combines the metric based on arcs and measurement of information content.

## 2. Organization of the database

In this step, we used a database of indexed video, based concepts (Ben Ammar, 2010). The following XML file (Figure 1) that describes the data in this database represents only video shots related to each concept. In this step, we aim to organize the data and facilitate access to them. This stage has two phases: data structuring and weighting of concepts.

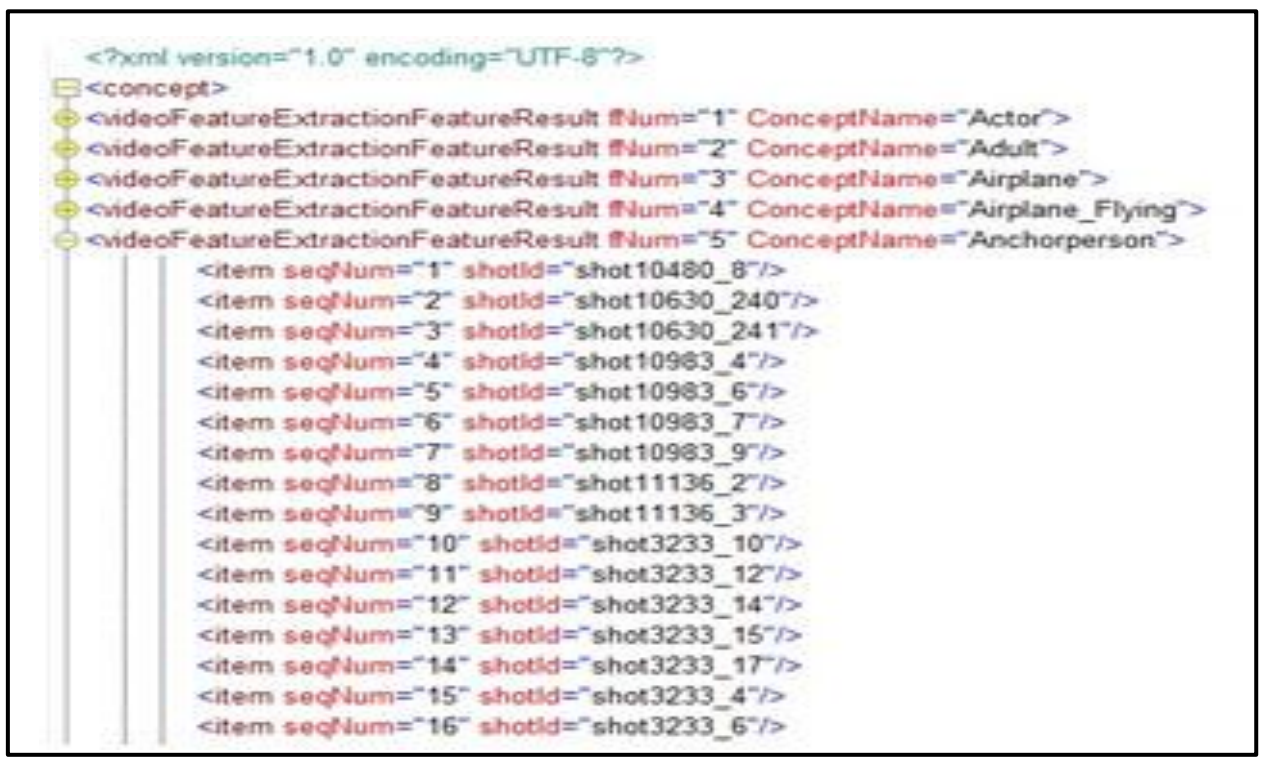

```
<?xml version="1.0" encoding="UTF-8"?>
<concept>
<videoFeatureExtractionFeatureResult fNum="1" ConceptName="Actor">
<videoFeatureExtractionFeatureResult fNum="2" ConceptName="Adult">
<videoFeatureExtractionFeatureResult fNum="3" ConceptName="Airplane">
<videoFeatureExtractionFeatureResult fNum="4" ConceptName="Airplane_Flying">
<videoFeatureExtractionFeatureResult fNum="5" ConceptName="Anchorperson">
    <item seqNum="1" shotId="shot10480_8"/>
    <item seqNum="2" shotId="shot10630_240"/>
    <item seqNum="3" shotId="shot10630_241"/>
    <item seqNum="4" shotId="shot10983_4"/>
    <item seqNum="5" shotId="shot10983_6"/>
    <item seqNum="6" shotId="shot10983_7"/>
    <item seqNum="7" shotId="shot10983_9"/>
    <item seqNum="8" shotId="shot11136_2"/>
    <item seqNum="9" shotId="shot11136_3"/>
    <item seqNum="10" shotId="shot3233_10"/>
    <item seqNum="11" shotId="shot3233_12"/>
    <item seqNum="12" shotId="shot3233_14"/>
    <item seqNum="13" shotId="shot3233_15"/>
    <item seqNum="14" shotId="shot3233_17"/>
    <item seqNum="15" shotId="shot3233_4"/>
    <item seqNum="16" shotId="shot3233_6"/>
```

**Figure 1 : XML file representing shots for each concept**

### 2.1. Structuring data

We structure our data according to three levels of abstraction: the contextual level, conceptual level and the level of raw data. Our idea is therefore to bring together all the data having common features and have a common concept. For example, we assume that:

- The features of a video (V1) are: "Car", "Driver",
- The features of a video (V2) are: "For", "Crowd", "Dogs",
- The features of a video (V3) are: "Dogs", "Cats".

The transition to another level of abstraction helps us to organize the data and to accelerate access. Similarly, we grouped semantically most similar concepts in the same context by adopting the work of (Ksibi, 2011) (Elleuch, 2011) for the extraction of concepts from context. We will study the same example to explain the transition to the third level:

- "Car" is a "Vehicle",
- "Driver" is a "Person",
- "Crowd" is a "Person",
- "Dogs" is an "Animal",
- "Cats" is an "Animal".

General representation of our database is as follows:

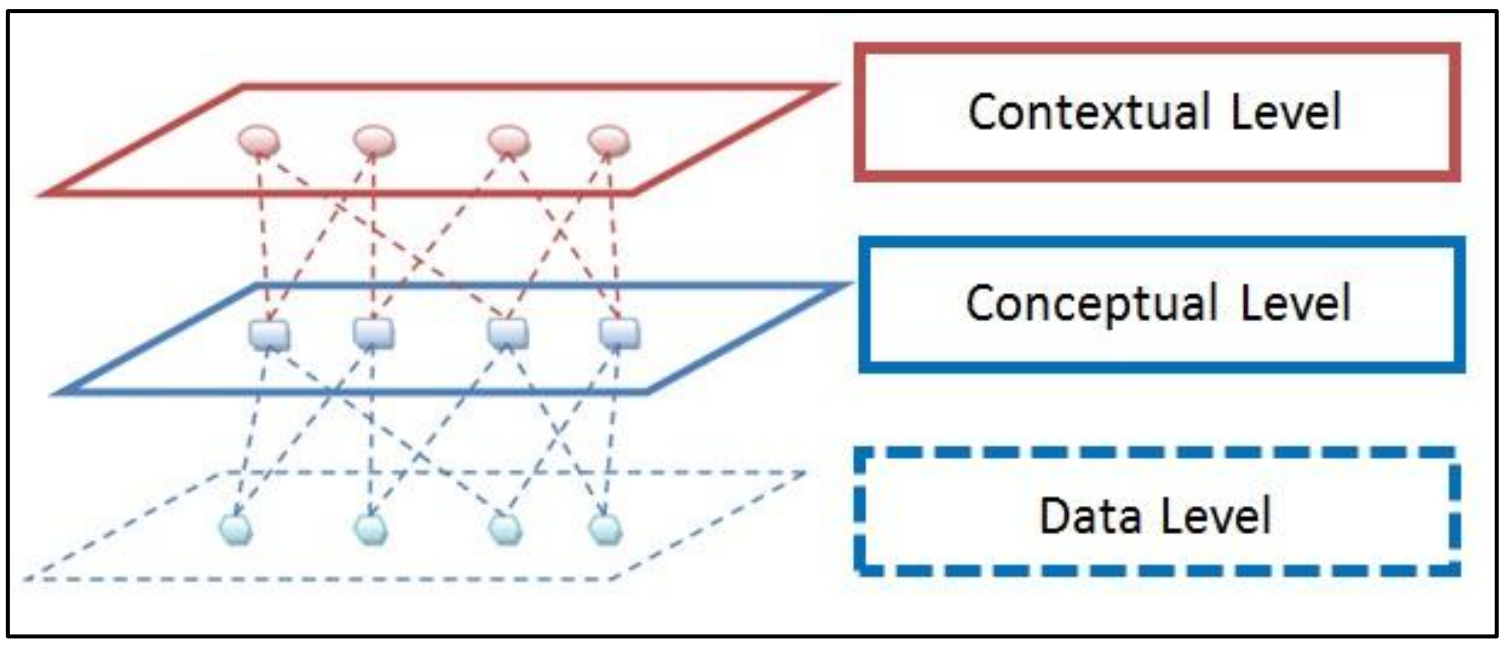

**Figure 2 : Data-Base structure**

## 2.2. Weighting of concepts

The weighting is one of the fundamental functions in information retrieval. We use an approach based on TF*IDF because it remains the most used by the information retrieval systems (Chung, 2008) (Nister,2006). The local weighting is used to measure the local representation of a concept.

$$tf(c_i, v_j) = P_1(c_i, v_j) \qquad (1)$$

Where

$$P_1(c_i, v_j) = \frac{nb_shots(c_i, v_j)}{n} \qquad (2)$$

$nb_shots(c_i, v_j)$ the number of shots includes the concept $c_i$ in a video $v_j$ and $n$ the total number of shots in the video.

$$tf(c_i, v_j) = P_2(c_i, v_j) \qquad (3)$$

The aggregate weighting of concepts. It depends on the document collection.
Where,

$$P_2(c_i, v_j) = \frac{nb_concepts_simi\ (c_i)}{nb_concepts(\ v_j) * \mathrm{N}} \qquad (4)$$

$nb_concepts_simi\ (c_i)$ is the number of concepts that are semantically similar to the concept $c_i$. $nb_{concepts(\ v_j)}$ is the number of concepts related to video $v_j$. And N is the number of sequences containing the concept $c_i$.
The combination of the formula (1) and the formula (2) provided the following formula:

$$P(c_i, v_j) = P_1 \times P_2 \qquad (5)$$

Otherwise,

$$P(c_i, v_j) = \left[\frac{nb_shots(c_i, v_j)}{n}\right] \times \left[\frac{nb_concepts_simi\ (c_i)}{nb_concepts(\ v_j) * \mathrm{N}}\right] \qquad (6)$$

# 3. Multilingual search (English/Arabic) using textual query

In this section we will detail two search sub-systems: A Multilingual sub-system Search (Arabic/English) based on text queries and another sub-queries based on images queries.
Usually, the easiest way for a user to express his request is a text that defines the content of videos which he want to see. The problem here is how to translate a textual query into concepts? To solve this problem we use the technique of query expansion by the call for ontology to help users to formulate their queries. Figure 3 explains the different stages of the research proposed system, based on text queries.

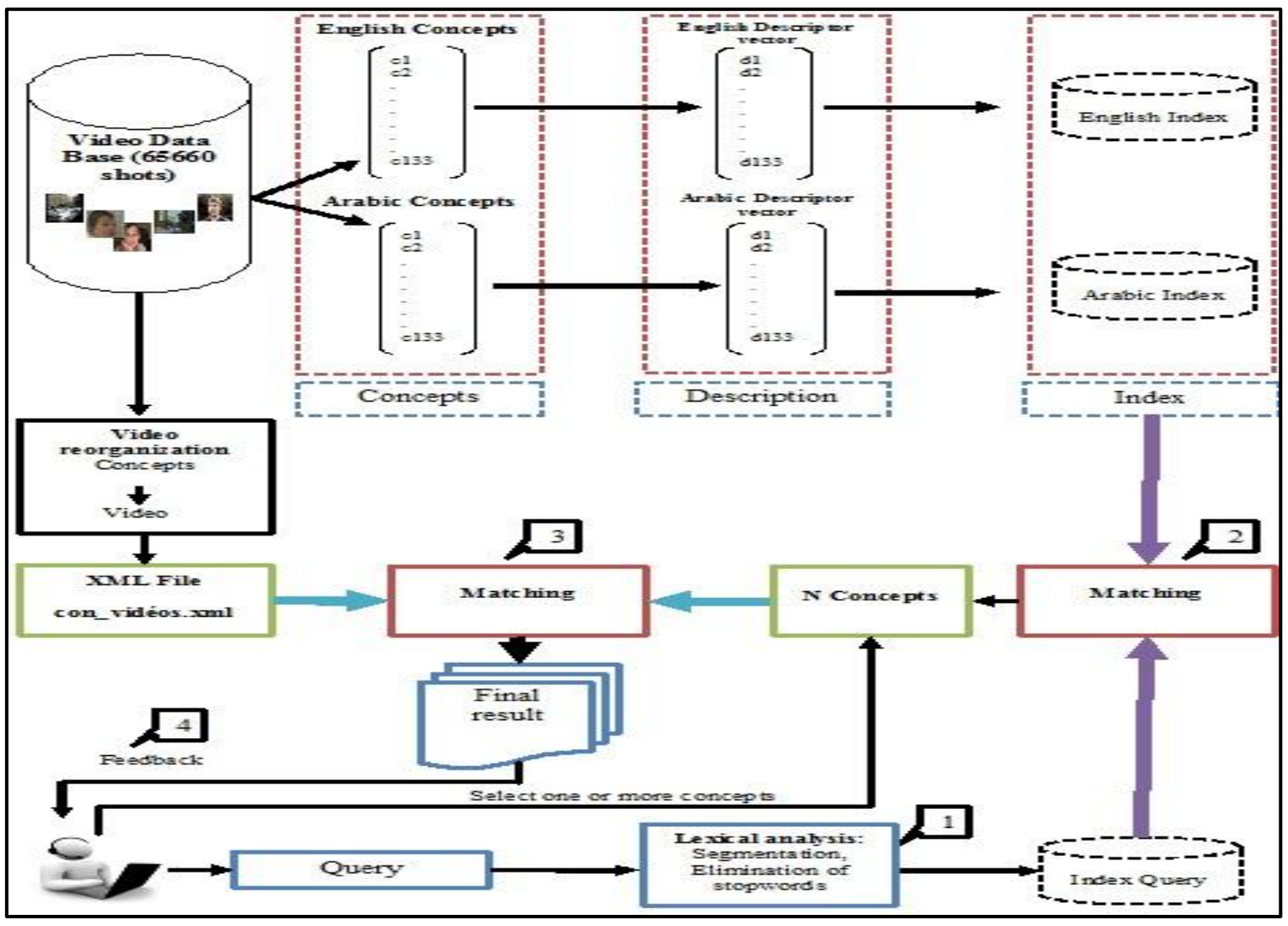

**Figure 3: Conceptual Architecture for research proposed system (textual query)**

The different steps are:

1. Indexing the query, by sharing equally the weights between terms that compose the query by removing stop words.
2. Make a comparison between the query terms and terms which are assigned to describe each concept among our database and returns concepts that have similar descriptors to the query. Extracting descriptive term for each concept is based on a specific description assigned to it.
3. In this step, the user operates to interactively choose the corresponding concepts.
4. Mapping between the request concepts and those of the whole collection is done using a vector model. This model incorporates a vector space representation which symbolizes documents or requests based indexing terms which compose them.

The advantage of this model resides in its simplicity. To calculate the degree of relevance of each video we use the cosine measure. This measure is the most used in the vector model.

$$simi(v_i, req) = \cos(\overrightarrow{v_i, req}) = \sum_{j=1}^{n} \frac{P_{c_j}(v_i) \times P_{c_j}(req)}{\sqrt{\sum_{k=1}^{n}\left(P_{c_k}(v_i)\right)^2} \times \sqrt{\sum_{k=1}^{n}\left(P_{c_k}(req)\right)^2}} \qquad (7)$$

Where, $req$ is the query, $v_i$ is a video of index i, $P_{c_j}(v_i)$ the weight of concept j of the video i and $P_{c_j}(req)$ the weight of the concept j of the query.

If the user not satisfied with the returned results, it can so improve the search with the technique of relevance feedback. So when a response is given by the system, the user can judge the returned videos to indicate those are relevant and those are irrelevant. Thus, the system recalculates and gives new results to the user.

## 4. Implementation and results

To properly evaluate our system, we use a database of 2700 and about 62,838 video keyframes and two XML file. The first describes the 130 concepts with their corresponding plans (Figure 4). The second XML file describes the relationship context-concept (Figure 5):

```xml
<?xml version="1.0" encoding="UTF-8"?>
<concept>
<videoFeatureExtractionFeatureResult fNum="1">
<item seqNum="1" shotId="shot11176_10"/>
<item seqNum="2" shotId="shot11176_11"/>
<item seqNum="3" shotId="shot11176_12"/>
          <item seqNum="4" shotId="shot11176_3"/>
          <item seqNum="5" shotId="shot11176_4"/>
          <item seqNum="6" shotId="shot11176_5"/>
          <item seqNum="7" shotId="shot11176_6"/>
          <item seqNum="8" shotId="shot11176_7"/>
          <item seqNum="9" shotId="shot11176_8"/>
          <item seqNum="10" shotId="shot11176_9"/>
          <item seqNum="11" shotId="shot11249_10"/>
          <item seqNum="12" shotId="shot11249_11"/>
          <item seqNum="13" shotId="shot11249_12"/>
          <item seqNum="14" shotId="shot11249_13"/>
          <item seqNum="15" shotId="shot11249_14"/>
          <item seqNum="16" shotId="shot11249_15"/>
          <item seqNum="17" shotId="shot11249_16"/>
          <item seqNum="18" shotId="shot11249_17"/>
          <item seqNum="19" shotId="shot11249_18"/>
          <item seqNum="20" shotId="shot11249_19"/>
          <item seqNum="21" shotId="shot11249_20"/>
          <item seqNum="22" shotId="shot11249_21"/>
          <item seqNum="23" shotId="shot11249_3"/>
          <item seqNum="24" shotId="shot11249_4"/>
          <item seqNum="25" shotId="shot11249_5"/>
```

**Figure 4. XML file representing the shots for each concept**

```xml
<?xml version="1.0" encoding="UTF-8"?>
 <contextes>


<Contexte Num="2" Name="Adult" Nbrconcept="6">
    <concept ConceptId="5" ConceptName="Anchorperson" Weight="0,6758"/>
    <concept ConceptId="11" ConceptName="Beards" Weight="0,4138"/>
    <concept ConceptId="34" ConceptName="Corporate_Leader" Weight="1"/>
    <concept ConceptId="83" ConceptName="News_Studio" Weight="0,7786"/>
    <concept ConceptId="86" ConceptName="Old_People" Weight="0,9816"/>
    <concept ConceptId="97" ConceptName="Reporters" Weight="0,8977"/>
</Contexte>

<Contexte Num="3" Name="Airplane" Nbrconcept="2">
    <concept ConceptId="4" ConceptName="Airplane_Flying"  Weight="1"/>
    <concept ConceptId="62" ConceptName="Helicopter_Hovering"  Weight="1"/>
</Contexte>


<Contexte Num="6" Name="Animal" Nbrconcept="5">
    <concept ConceptId="14" ConceptName="Birds" Weight="1"/>
    <concept ConceptId="23" ConceptName="Cats" Weight="1"/>
    <concept ConceptId="36" ConceptName="Cows" Weight="1"/>
    <concept ConceptId="43" ConceptName="Dogs" Weight="1"/>
    <concept ConceptId="64" ConceptName="Horse" Weight="1"/>
</Contexte>
```

**Figure 5. XML file representing the different context**

A user will launch an initial request $\boldsymbol{Q_0}$ then do three iterations from this query. Among the first 60 videos returned by the system, the user will judge "$m$" relevant videos and "$n$" irrelevant videos. We use the recall which determines the possibility of the system to retrieve all relevant documents. It is the ratio between the number of retrieved documents and number of documents in the database for a given query. We also use the precision, which determines, for a given query, the system's ability to present only relevant documents. It is the ratio between the number of retrieved documents and number of documents returned (relevant and irrelevant).

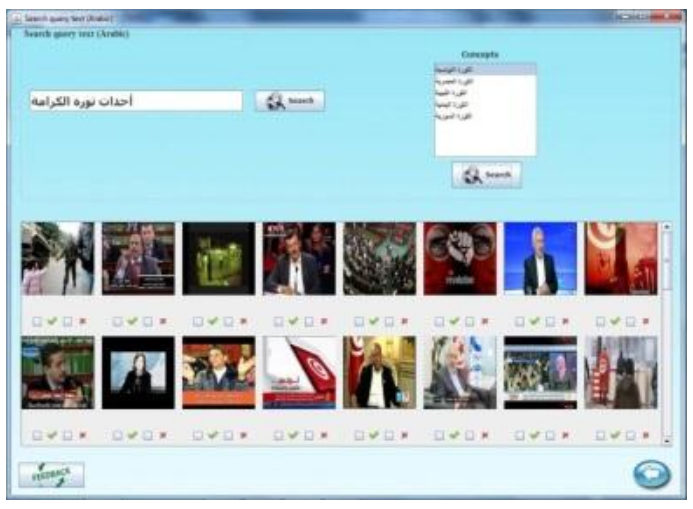

**Figure 6. search result of Arabic textual query "أحداث ثورة الكرامة"**

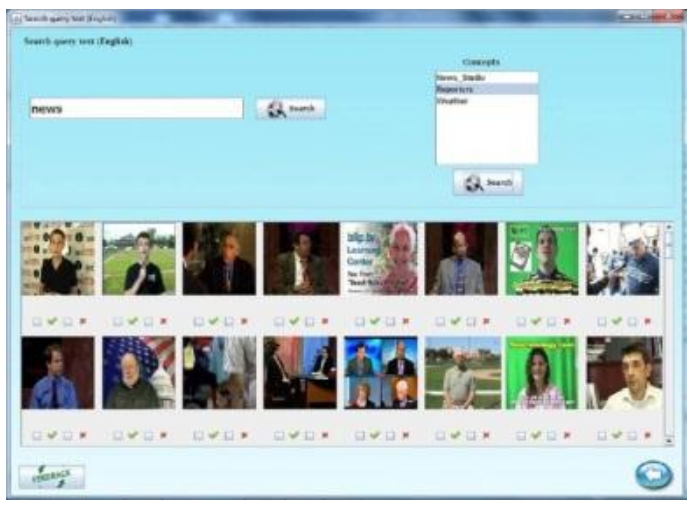

**Figure 7. search result of English textual query "news"**

By applying the precision and recall, we get the following curve:

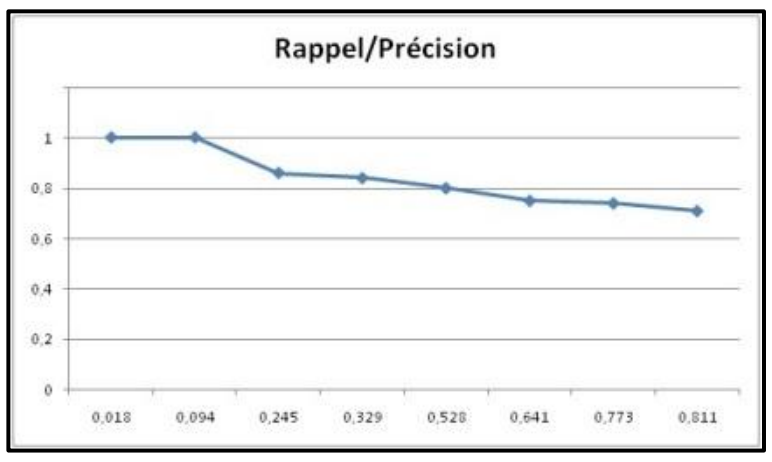

**Figure 8. Recall and precision corresponding to the original query "$Q_0$"**

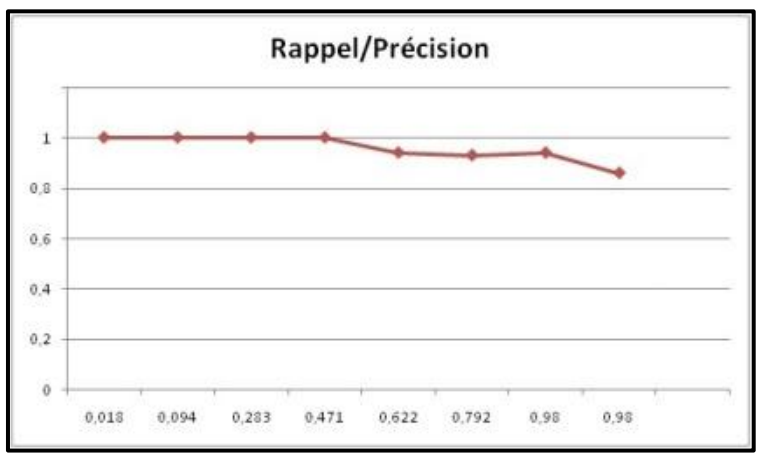

**Figure 9. Recall and precision corresponding to the query "$Q_1$"**

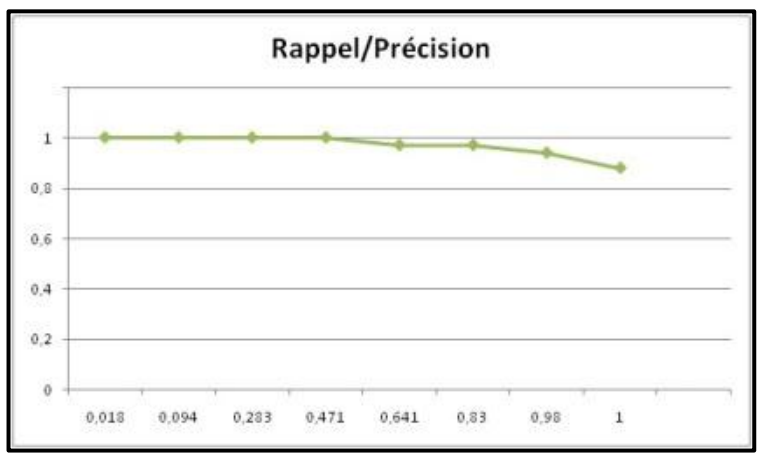

**Figure 10. Recall and precision corresponding to the query "$Q_2$"**

According to Figure 8, we note that the accuracy begins to decrease from the point (0.094). This decrease is explained by the presence of irrelevant videos in the result from the point (0.094). According to precision values at different points, we can judge that the precision at the level of query "$Q_0$" is acceptable.

Now, we perform the same process to calculate the recall and precision for the new query "$Q_1$" from the initial query "$Q_0$" using the relevance feedback formula (8). We get the following result at figure 9. Similarly, we will repeat the same process to calculate the recall and precision for a new query "$Q_2$" from previous query "$Q_1$". We get the following result at figure 10.

$$PQ_i = P_{initial} + P_{fb_i} \pm \propto \qquad (8)$$

Where, $Q_0 = P_{initial} + P_{fb_0}$ , $P_{fb_i} = PQ_{i-1}$ , $P_{fb_0} = 0$ and $\propto$ $\pm 0.2$

According to Figure 10, we see stability in values clarification related to the first four points. This means that the relevant videos are located at the beginning of result. For the other points, the precision values are higher than 0.88 and later are superior to those of the query "$\boldsymbol{Q_1}$".

## 5. Conclusion

In this paper, we validated our proposal system to search video from a corpus of audiovisual documents (TRECVID 2010) characterized by its size and importance of its content heterogeneity. We have developed a system of video search easy access to desired video. Thus, our system provides a method of relevance feedback to improve performance.